%% file: arxiv.tex
\documentclass[3p,review,sort&compress]{elsarticle}

\usepackage[svgnames]{xcolor}
\usepackage{graphicx}      
\usepackage{natbib}
\usepackage{amsmath}
\usepackage{bm}
\usepackage{amssymb}
\usepackage{threeparttable}
\usepackage{multirow}
\usepackage{ucs}
\usepackage[utf8x]{inputenc}
\usepackage[T1]{fontenc}
\usepackage{engord}
\usepackage[subrefformat=parens,labelformat=parens]{subfig}
\usepackage[inline]{enumitem}
\usepackage{booktabs}
\usepackage{mathrsfs}
\usepackage[colorinlistoftodos]{todonotes}
\usepackage[acronym]{glossaries}
\usepackage{subfig}
\usepackage{xfrac}
\usepackage{engord}
\usepackage{enumitem}
\usepackage[version=4]{mhchem}
\usepackage[section]{placeins}
\usepackage{siunitx}
\usepackage{arydshln}

\usepackage{bm}
\usepackage{algorithm}
\usepackage{algorithmicx}
\usepackage{algpseudocode}
\usepackage{xcolor}
\usepackage{threeparttable}

\usepackage{tabularx} 
\usetikzlibrary{positioning,shapes,shadows,arrows,decorations.pathreplacing}
\usetikzlibrary{arrows, decorations.markings}
\usetikzlibrary{matrix,decorations.pathmorphing,positioning}
\usetikzlibrary{fit}
\usetikzlibrary{calc}
\usetikzlibrary{arrows}

\usepackage{glossaries}
\newacronym{pes}{PES}{public employment service}
\newacronym{kl}{KL}{Kullback-Leibler}
\newacronym{vb}{VB}{variational Bayes}
\newacronym{fos}{FOS}{fractional order system}
\newacronym{svi}{SVI}{stochastic variational inference}
\newacronym{elbo}{ELBO}{evidence lower bound}
\newacronym{eis}{EIS}{electrochemical impedance spectroscopy}
\newacronym{mcmc}{MCMC}{Markov chain Monte Carlo}
\newacronym{sofc}{SOFC}{solid-oxide fuel cell}
\newacronym{ecm}{ECM}{equivalent circuit model}
\newacronym{drt}{DRT}{distribution of relaxation times}
\newacronym{adam}{ADAM}{adaptive moment estimation}
\newacronym{rnn}{RNN}{recurrent neural networks}
\newacronym{ann}{ANN}{artificial neural network}

\usepackage[normalem]{ulem}
\newcommand{\mrev}[2]{#2 }
\newcommand{\mrevsec}[2]{#2 }

\usepackage{tikz}
\usetikzlibrary{bayesnet}
\usetikzlibrary{decorations.pathreplacing}

\usepackage{url}

\begin{document}
\engordraisetrue
\begin{frontmatter}
	\title{Variational Bayes survival analysis for unemployment modelling}

\author[1]{Pavle Boškoski\corref{cor1}}
\ead{pavle.boskoski@ijs.si}

\author[1]{Matija Perne}

\author[3]{Martina Rameša}

\author[1,2]{Biljana Mileva Boshkoska}


\address[1]{Jožef Stefan Institute, Jamova cesta 39, 1000 Ljubljana, Slovenia}

\address[2]{Faculty of information sciences in Novo mesto, Ljubljanska cesta 31a, 8000 Novo mesto, Slovenia}

\address[3]{Employment service of Slovenia, Rožna dolina, Cesta IX/6, 1000 Ljubljana, Slovenia}

	\cortext[cor1]{Corresponding author.}

\begin{abstract} 


Mathematical modelling of unemployment
dynamics attempts to predict the probability of a job seeker finding a job as a function of time.
This is typically achieved by using information in unemployment records.
These records are right censored, making survival analysis a suitable approach for parameter estimation.
The proposed model uses a deep \gls{ann} as a non-linear hazard function.
Through embedding, high-cardinality categorical features are analysed efficiently.
The posterior distribution of the \gls{ann} parameters are estimated using a variational Bayes method.
The model is evaluated on a time-to-employment data set spanning from 2011 to 2020 provided by the Slovenian \glsdesc{pes}.
It is used to determine the employment probability over time for each individual on the record.
Similar models could be applied to other questions with multi-dimensional, high-cardinality categorical data including censored records.
Such data is often encountered in personal records, for example in medical records.
\end{abstract}

	
\end{frontmatter}
\glsresetall

\input{introduction.tex}
\input{survival.tex}
\input{vb.tex}

\input{surv_vb.tex}
\input{results.tex}
\input{conclusion.tex}

\appendix
\input{appendix.tex}

\section*{Acknowledgements}
The authors acknowledge the research core funding No.\ P2-0001 and P1-0383 that were financially supported by the Slovenian Research Agency.
The authors also acknowledge the funding received from the European Union's Horizon 2020 research and innovation programme project HECAT under grant agreement No.\ 870702 .

\bibliographystyle{elsarticle-num-names}
\bibliography{references}

\end{document}

%% file: introduction.tex

\section{Introduction}
A reliable time-to-employment estimate is a valuable piece of information both for the job seekers and for the \gls{pes} employment counsellors.
Perceived employability has important effects on job seekers~\cite{yizhong2017employability} so improving its accuracy may be beneficial.
Long-term unemployment has been identified as having a significant scarring effect on society and the economy and, more importantly, on the health of the unemployed persons~\cite{10.2307/798307,VIRTANEN201346,10.1093/eurpub/cku005}.
As a result, the \gls{pes} counsellors want to focus their attention on those job seekers that need their help.
A time-to-employment prediction can help them identify the ones that do not need \gls{pes} resources as they will get employed soon regardless of the interventions.
Algorithmic tools predicting the future of particular job seekers are therefore needed to support the decision making process.

The field of creating such tools is very active.
One can track such efforts for over 20 years~\cite{Grundy2015}. 
The methods used can be roughly separated into four groups.
The first and most numerous group consists of approaches based on logit/probit models~\cite{Riipinen2011, Wijnhoven2014, O_CONNELL_2009, Loxha2014}.
The second group consists of so-called in/out models, whose goal is to estimate the probabilities of entering and exiting the labour market~\cite{Sengul2017, Shimer2012}.
The biggest limitation with these two groups is their inability to incorporate a large number of high-cardinality discrete data.
The data used are typically yes/no questionaries, hence the limited efficiency.
The third group includes machine learning approaches~\cite{10.1111/ijsw.12088}.
These approaches, on the other hand, are capable of handling vast amounts of heterogeneous data.
However, the results are almost always point estimates, and the lack of uncertainty assessment can have significant negative practical consequences.
Finally, there are the emerging approaches based on labour flow networks with the goal of modelling the labour market as a dynamic system~\cite{Park2019,Lopez2015}.
These models require high quality micro-data that usually have restricted access.

The biggest issue when modelling the labour market, or in other terms, modelling the dynamics of the unemployed part of the population, is discovering the influencing forces.
It is well known that it is impossible to infer the overall system dynamics through modelling the behaviour of each individual ``actor'' in a system~\cite{Anderson1972}.
The same is shown to be valid for sociological systems~\cite{Kittel2006}.
Therefore, it is important to take various interdependencies between different societal levels into account~\cite{Erlinghagen2019}.

As the modern society evolves quickly,
old information describing socio-economical dynamics bears little importance for current or future behaviour.
The amount of historical data that can be effectually used for deriving the current and future dynamics is therefore limited.
This inevitably leads to censored data points~\cite{Huang2015,Robins1995}.
The concept of censoring is important because ignoring or otherwise mistreating the censored cases might lead to false conclusions~\cite{doi:10.1080/03610918.2015.1056355}.

Survival analysis is capable of properly handling cases of censored time-to-event data~\cite{Ibrahim2001}.
The most prominent example is the Cox proportional hazards model~\cite{Cox1972}.
With a linear risk function in the Cox model, the complete likelihood can be derived in a closed form.
However, this limits the applicability of the model for systems with more complex and non-linear dynamics.

The issue with more complex functions is that the solution of the Cox model becomes intractable.
An alternative is to employ Markov chain Monte Carlo approaches~\cite{Ibrahim2001}.
However, there are two main limitations of such approaches: the immense computational load for multidimensional data and the presence of discrete/categorical data.
The latter issue is usually resolved through so-called one-hot encoding, which for large cardinality covariates causes an explosion of model parameters.

A logical step forward in survival analysis is the use of machine learning~\cite{Faraggi1995,KOVALEV2020106164}.
These techniques are capable of handling mixtures of continuous and categorical data through the concept of embedding~\cite{Guo2016}. 
\citet{Kvamme2019} have derived the loss function for a survival analysis model using a deep neural network as its core.
There have been several recent results that build upon Cox proportional hazards models~\cite{Katzman:2018aa,8681104,CoxNet}.
Additionally, there are proposals in which parametric survival models employ \gls{rnn} in order to predict the empirical probability distribution of future events~\cite{Giunchiglia2018, martinsson:Thesis:2016}.
These models have undisputed applicability with one drawback.
They provide only point estimates of the posterior distributions of the survival/hazard functions.

\mrev{Comment 1.1.}
{
A point estimate only provides a partial view of the results. 
For example, in a classification problem, this would mean providing the most probable class without the assessment of the probability that a certain entity belongs to the selected class.
The cases where the confidence in the classification is overwhelming (for instance 90\%)
and the cases where the confidence is low (for instance 51\%) would result in the same output.
However, these cases must be handled very differently in any subsequent decision steps.
In the presented case of survival analysis, where the underlying phenomenon is inherently stochastic, it is clear that maximum likelihood estimates do not provide the whole picture.
}

A method harnessing the power of obtaining complete posterior distributions, like with Markov chain Monte Carlo approaches, while preserving the efficiency of the machine learning methods, is the \gls{vb} method~\cite{Smidl2006}. 
It has been successfully applied in various fields, such as Gaussian processes modelling~\cite{Bui2016, Hensman2014}, deep generative models~\cite{Rezende2014}, compressed sensing~\cite{Yang2013,Oikonomou2019}, hidden Markov models~\cite{Gruhl2016, Panousis2020}, reinforcement learning and control~\cite{Levine2018}, etc.
It is possible to define a survival model with an arbitrary risk function that is implemented as an \gls{ann} and
estimate the posterior distribution of the \gls{ann} model parameters despite the large number of parameters.

All these properties come to use when building survival models using personal records, for instance medical, \gls{pes} data, or similar.
The records tend to be multi-dimensional and usually have high-cardinality categorical data~\cite{PIVOVAROV201424,POLSTERL20161,doi:10.1002/int.21825}.
Such data are an ideal candidate for harnessing the capability of the \gls{vb}-based survival method.

The proposed model estimating time-to-employment is a survival model with an \gls{ann} used as a non-linear hazard function.
\mrev{Comment 2.1.}{
Half of the covariates are categorical, some of them with cardinality of more than 2000.
Using one-hot encoding, a traditional approach for survival analysis, would lead to an ``explosion'' of the number of parameters, thus harming the efficiency of the parameter estimation process.
In such a case, the maximum-likelihood estimators became intractable. 

	Therefore, instead of limiting the analysis to some form of linear embedding, we opt for a completely free choice of hazard rate function.
	Consequently we are able to harness the whole potential of the current methods addressing the analysis of high-cardinality categorical data.
	}

\mrev{Comment 1.1.}
{
The original contribution of this work is the proposed use of \gls{vb} method in a survival analysis with \gls{ann} as the hazard rate. 
It may serve as a generally applicable algorithm for survival analysis.
It enables seamless integration of various approximations of the hazard rate or the form of the survival function.
Furthermore, the changes of the underlying error distribution or the distribution of survival times can be easily varied.
}

A basic overview of survival analysis is given in Section~\ref{sec:survival}.
Section~\ref{sec:vb} presents the \glsdesc{vb} approach.
The proposed \gls{vb}-based survival analysis solution is presented in Section~\ref{sec:surv_vb}.
The application on unemployment records is described in Section~\ref{sec:pes}.
The results are presented in Section~\ref{sec:res} and discussed in Section~\ref{sec:dis}.


%% file: survival.tex

\section{Survival analysis}
\label{sec:survival}
Survival analysis is the study of time-to-event data that 
initially focused on estimation of lifespans. 
The majority of the developed methods investigate continuous-time models~\cite{Klein2003}, but
there are also developments on discrete-time approaches~\cite{Tutz2016}.

Survival analysis explores the probability distribution of an event over time.
The probability that an event occurs at time $T$ before a certain time $t$ can be written as
\begin{equation}
\Pr(T \le t)=\int_0^t f(s) ds = F(t),
\label{eq:prob_t}
\end{equation}
where the functions $f(t)$ and $F(t)$ are the probability density and the cumulative probability function, respectively.
The opposite case, i.e. the probability that the time of occurrence $T$ will be after a certain time $t$, is
\begin{equation}
\Pr( T > t) = S(t) = 1-F(t).
\end{equation}
The function $S(t)$ is called the survival function.
The hazard rate, or hazard function, $\lambda(t)$ is defined as the event rate at time $t$ if the event has not occurred up until $t$ and is expressed as
\begin{equation}
\lambda(t) = \lim_{\Delta t \to 0}\frac{1}{\Delta t} \Pr(t\le T < t+\Delta T|T \ge t) = \frac{f(t)}{S(t)}.
\end{equation}
The survival function can be obtained from the hazard function as
\begin{equation}
S(t) = \exp\left[ -\Lambda(t) \right],\quad \Lambda(t) = \int_0^t \lambda(s) ds.
\end{equation}

Survival models are categorised based on the type of the hazard rate.
Often, the multiplicative hazard function is used~\cite{Klein2003},
\begin{equation}
\lambda(t|\mathbf{x}) = \lambda_0(t) e^{h(\mathbf{x})}.
\label{eq:lam}
\end{equation}
Its two components are the baseline hazard $\lambda_0(t)$ and a risk function $h(\mathbf{x})$, which depends on directly measurable covariates $\mathbf{x}$ that are also known as explanatory variables or features.
In the simplest form, i.e. Cox proportional hazards models, the risk function is linear, $h_{\beta}(\mathbf{x}) = \beta^T \mathbf{x}$, and needs no constant term as it is included in the baseline term $\lambda_0(t)$.
The parameter set $\beta$ is identified from the data.
This is achieved using various estimation approaches, the most popular being the Kaplan-Meier estimator~\cite{10.2307/2281868}.
For the case of non-linear models, the risk function $h(\mathbf{x})$ can be an arbitrary real function.
In such cases, various forms of \gls{ann}s have been applied~\cite{Kvamme2019,Katzman:2018aa,8681104}.

Another possibility is the use of accelerated failure-time models~\cite{Klein2003}, where the logarithm of the survival time $y$ is approximated with linear regression,
\begin{equation}
\log T = y = \beta^T\mathbf{x} + \sigma W.
\label{eq:lin_acc}
\end{equation}
The probability distribution of the error 
 is altered using regression coefficients $\beta$, covariates $\mathbf{x}$, and scale factor $\sigma$ to obtain the probability distribution of the event time $T$.
The choice of the probability distribution of $W$ directly defines the probability distribution of the survival time $T$.
Typical pairs are listed in \tablename~\ref{tab:dist}.

The model can be extended by replacing the linear function $\beta^T\mathbf{x}$ with an arbitrary real function of the covariates $h_{\mathbf{z}}(\mathbf{x})$,
where $\mathbf{z}$ is the set of the parameters of the non-linear function.
 When the model~\eqref{eq:lin_acc} is extended this way, the equation reads
\begin{equation}
y = h_{\mathbf{z}}(\mathbf{x}) + \sigma W.
\label{eq:non_lin}
\end{equation}
It has been done with the output of an \gls{ann} used as $h_{\mathbf{z}}(\mathbf{x})$ \cite{XIANG2000243,Faraggi1995,Kvamme2019,pmlr-v56-Ranganath16}.

\begin{table}[bt]
\centering
\caption{Error distribution and corresponding survival time distributions.}
\begin{tabular}{lr}
\toprule
Log-Error distribution & Survival time distribution \\
\midrule
Normal distribution & Log-normal \\
Gumbel (Extreme value) & Weibull \\
Logistic & Log-logistic \\
\bottomrule  
\end{tabular}
\label{tab:dist}
\end{table}

\subsection{Censoring}
Two particularities of time-to-event data sets that complicate their processing are censoring and truncation~\cite{Klein2003}.
Censoring happens when the time of event may only be known to be within a particular time interval. 
Truncation occurs when the cases with event times outside of the observation period are not observed.
In the studied example, there is no truncation and only right censoring, also known as Type I censoring. That is, all the subjects are observed, the timing of the event is known if it occurs prior to the end of the observation time, and the exact timing is unknown for the events that occur later.

Censoring might lead to false conclusions if it is not accounted for properly~\cite{doi:10.1080/03610918.2015.1056355}.
The distinction between survival analysis and simple regression is in the handling of censored data.


\subsection{Addressed limitations of typical survival models}


The most common survival analysis models are of the types \eqref{eq:lam} and \eqref{eq:non_lin}
and limited to such functions $h(\mathbf{x})$ or $h_{\mathbf{z}}(\mathbf{x})$ 
that 
the likelihood function exists in closed form~\cite{Klein2003}.
However, the modelling benefits from the use of more general functions
that are capable of handling non-linear relationships between the covariates $\mathbf{x}$.
For those, the evidence $p(x)$ in equation \eqref{eq:BT} is intractable.
Solving the equation \eqref{eq:BT} in order to estimate the posterior probability distribution $p(\theta|x)$ of the model parameters $\theta$ thus requires the use of an approximation.

We choose to use an \gls{ann} in place of $h_{\mathbf{z}}(\mathbf{x})$ in equation \eqref{eq:non_lin} and to estimate the posterior probability distribution $p(\theta|x)$ using the \gls{vb} method \cite{Smidl2006}.
Another issue is the use of 
high-cardinality discrete covariates, necessitating some form of dimension embedding.
Such properties are becoming typical in medical records~\cite{PIVOVAROV201424,POLSTERL20161,doi:10.1002/int.21825} and are also present in this study.

%% file: vb.tex
\section{Variational Bayes method}
\label{sec:vb}
When performing a stochastic analysis, such as survival analysis, the inference process of estimating the model parameters relies on the Bayes' rule.
For a set of observations $x$,
generated by a system with parameters $\theta$, the Bayes' rule reads
\begin{equation}\label{eq:BT}
\underbrace{p(\theta|x)}_{\text{Posterior}} = \frac{\overbrace{p(x|\theta)}^{\text{Likelihood}}\overbrace{p(\theta)}^{\text{Prior}}}{\underbrace{p(x)}_{\text{Evidence}}}.
\end{equation}
The likelihood $p(x|\theta)$ is typically known because it is prescribed by the model structure.
The prior $p(\theta)$ is typically chosen.
The biggest obstacle in computing the posterior probability distribution $p(\theta|x)$ is the evidence $p(x)$.
It is the solution of the equation
\begin{equation}
p(x) = \int_{\theta}p(x|\theta)p(\theta)d\theta,
\label{eq:evidence}
\end{equation}
which in most cases cannot be obtained in a closed form.
For multi-dimensional cases, even Monte Carlo integration becomes impractical due to the immense computational load.
The \gls{vb} method provides an approximative solution to this problem.
	
The idea of the \gls{vb} method is to sufficiently closely approximate the true posterior $p(\theta|x)$ with an approximative probability distribution $q_{\omega^*}(\theta)$, referred to as \emph{variational distribution}. It belongs to a \emph{variational family} of the functions $q_{\omega}(\theta) \in \mathcal{Q}$, $\omega\in\Omega$, where $\Omega$ is the set of all possible values of the latent parameters $\omega$.
A typical variational family is
the mean-field variational family~\cite{Blei2017} in which the parameters $\theta$ are mutually independent.

The optimal values of the latent parameters $\omega^*$ are obtained by minimising the \gls{kl} divergence $KL(q_{\omega}(\theta)||p(\theta|x))$ between the true posterior and the variational distribution.
The optimisation problem
\begin{equation}
\omega^*=\underset{\omega\in\Omega}{\arg\min}\gls{kl}(q_{\omega}(\theta)||p(\theta|x))
\label{eq:optim}
\end{equation}
is solved.

Since the true posterior is unknown, calculating the \gls{kl} divergence requires a minor rearrangement.
As shown by \citet{Smidl2006}, $KL(q_{\omega}(\theta)||p(\theta|x))$ can be written as 
\begin{equation}\label{eq:kl}
\begin{split}
KL(q_{\omega}(\theta)||p(\theta|x)) &=\mathbb{E}_q\left[ \log \frac{q_{\omega}(\theta)}{p(\theta|x)}\right]\\
& = \mathbb{E}_q\left[\log q_{\omega}(\theta)\right] - \mathbb{E}_q\left[ \log p(\theta|x)\right]\\
& = \mathbb{E}_q\left[ \log q_{\omega}(\theta) \right]- \mathbb{E}_q \left[\log p(x,\theta) - \log p(x)\right]\\
& = \mathbb{E}_q [\log q_{\omega}(\theta) - \log p(x,\theta)] + \log p(x)\\
& = - \underbrace{\mathbb{E}_q[\log p(x,\theta) - \log q_{\omega}(\theta)]}_{\text{ELBO}} + \log p(x)
\end{split}.
\end{equation}
The first term of the final expression is known as \gls{elbo} and maximising it results in minimising the \gls{kl} divergence between the variational distribution and the true posterior.
The second term, $\log p(x)$, is constant and is therefore not a part of the optimisation process.

\mrev{Comments 1.3 and first part of 1.6. The convergence on the real data is addressed in Section~\ref{sec:stopping}.}
{
Generally, the criterion \eqref{eq:kl} is not convex and no optimisation algorithm guarantees convergence to a global extreme.
Several optimisation algorithms have been used for this problem, such as coordinate ascent variational inference and conjugate models~\cite{Blei2017}, stochastic variational inference~\cite{hoffman2013stochastic}, black-box variational inference~\cite{ranganath2013black} and partially the automatic differentiation variational inference~\cite{kucukelbir2016automatic}.

In the presented case, the optimisation of the \gls{elbo} loss function is performed using the stochastic variational inference with gradients of loss function calculated following the black-box variational inference~\cite{ranganath2013black}.
According to \citeauthor{hoffman2013stochastic}~\cite[Eq. (23)]{hoffman2013stochastic}, this iterative algorithm converges to the optimal parameters if the objective function is convex or to a local optimum otherwise.
}

\begin{figure}[h]
\centering
\includegraphics{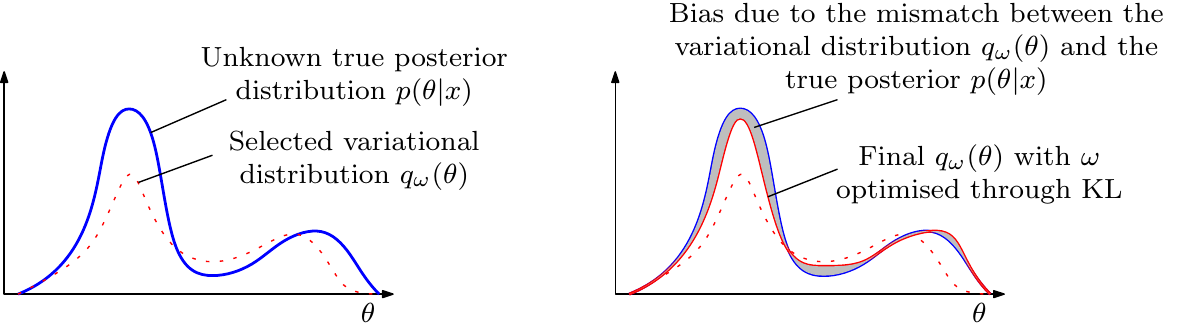}
\caption{Optimisation process of finding the closest variational distribution $q_{\omega}(\theta)$ over the set of latent variables $\omega$.}
\label{fig:vi}
\end{figure}

Having an approximative variational distribution instead of the true posterior distribution introduces an inherent bias which depends on the variational family used.
The selection of the variational family $\mathcal{Q}$  is thus not an ad-hoc decision but is based on
prior knowledge such as empirical observations or experts' knowledge. 
In spite of the inherent bias, the \gls{vb} method is justified by the substantial increase in the computational efficiency compared to the alternatives such as Monte Carlo integration.
In our analysis, ADAM optimiser~\cite{Kingma2014,Reddi2019}, implemented as a part of the \texttt{PyTorch} package~\cite{NEURIPS2019_9015}, is used for optimisation~\eqref{eq:optim} through Pyro~\cite{bingham2018pyro}.
The overall idea is schematically presented in \figurename~\ref{fig:vi}.

%% file: surv_vb.tex
\section{Variational Bayes method in survival analysis}
\label{sec:surv_vb}


The use of the model of the form \eqref{eq:non_lin} with an \gls{ann} as $h_{\mathbf{z}}(\mathbf{x})$ results in the integral \eqref{eq:evidence} that cannot be calculated analytically. Obtaining the posterior probability density function $p(\theta|x)$ thus requires an approximation.
\glspl{ann} typically have hundreds of parameters, precluding even the use of Monte Carlo methods.
Another option are Gaussian processes that are shown to be equivalent to a fully connected \gls{ann} with an infinite number of hidden units in each layer~\cite{Lee_2018}.
A variational inference approach provides a computationally efficient solution using \gls{elbo}~\eqref{eq:kl} as the loss function.
With the \gls{vb} method, one has to select the prior distribution $p(\theta)$ and the variational family of the posterior distribution $\mathcal{Q}$.
This is followed by optimisation \eqref{eq:optim} using the \gls{elbo} loss function~\eqref{eq:kl}, resulting in the variational distribution latent parameters $\omega^*$.
In the most general form of the \gls{vb} method, every model parameter may be treated as a stochastic one.


We implement the function $h_{\mathbf{z}}(\mathbf{x})$ as an \gls{ann}.
The model parameters are assembled as $\theta = [\sigma, \mathbf{z}]$, $\mathbf{z}] = [\mu_1,\ldots,\mu_K$.
We use a mean-field variational family, so the probability distribution $q_{\omega}(\theta)$ can be expressed as
\begin{equation}
q_{\omega}(\theta) = q_{0}(\sigma) \cdot \prod_{k=1}^K q_k(z_k),
\label{eq:meanfield}
\end{equation}
where $K$ is the number of the \gls{ann} parameters.
For the variational distribution, we use a half-normal distribution for the factor $q_{0}(\sigma)$ and normal distribution for every $q_k(z_k)$.
The vector of latent parameters $\omega$ is structured as $\omega=[\sigma_{\sigma},\omega_\mathbf{z}]$, $\omega_\mathbf{z}=[\mu_1,\ldots,\mu_K,\sigma_1,\ldots,\sigma_K]$.
The latent parameter $\sigma_{\sigma}$ is used in $q_{0}(\sigma) = \frac{2}{\sigma_{\sigma}}\phi\left(\frac{\sigma}{\sigma_{\sigma}}\right)$ for $\sigma \geq 0$ and the latent parameters $\omega_\mathbf{z}$ define $q_k(z_k) = \frac{1}{\sigma_k} \phi\left(\frac{z_k-\mu_k}{\sigma_k}\right)$, where $\phi$ is the probability density function of the standard normal probability distribution.

\begin{figure}[h]
\centering
\begin{tikzpicture}

\newcommand{\hgatep}[6]{ %
  \node[wrap=#2] (#1-top) {}; %
  \node[wrap=#4] (#1-bottom) {}; %
  \node[gate=(#1-top)(#1-bottom)] (#1) {}; %
  \node[caption, above left=of #1.east ] (#1-top-caption)
  {#3}; %
  \node[caption, below left=of #1.east ] (#1-bottom-caption)
  {#5}; %
  \draw [-, dashed] (#1.west) -- (#1.east); %
  \foreach \x in {#6} { %
    \draw [-*,thick] (\x) -- (#1); %
  } ;%
}


  
    \factor    {y}      {below:$W$} {} {} ; %
    \factor[left=of y,xshift=-1cm] {h_nn} {below:$h$} {} {} ;
    
    \factor[left=of h_nn, xshift=-2.cm]    {nn_params}  {above:$q_k(z_k)$} {} {};
    
    \node[latent, left=of nn_params, xshift=-.5cm]    (var_w)  {$\omega_k$};
    \node[latent, above=of var_w,yshift=-.0cm]    (wtheta)  {$\sigma_{\sigma}$}; %
    \factor[right=of wtheta, xshift=3cm]    {sigma}  {above:$q_0(\sigma)$} {} {};
    \node[obs, below left= of h_nn]	(x) {$\mathbf{x}^{(i)}$} ;
    
    \node[obs, right=of y, xshift=4cm]                   (trun)      {$c^{(i)}$} ; %
    \factor[below right=of y, xshift=1cm,yshift=-.18cm] {trun_obs} {below:1-CDF$_W$} {} {};
    
    \node[latent, right=of y,xshift=2cm, yshift=2.1cm,inner sep=0pt,minimum size=4pt] (leg1) {};
    \node[obs, below=of leg1,inner sep=0pt,minimum size=4pt, yshift=.8cm] (leg2) {};
    \node[right of=leg1, xshift=-.9cm, align=left,inner sep=0pt,anchor=west] {\tiny Latent variable};
    \node[right of=leg2, xshift=-.9cm, align=left,inner sep=0pt,anchor=west] {\tiny Observed variable};

    \factor[right=of trun_obs,xshift=1cm] {bern} {below:Bernoulli} {}{}; 
    \node[const,right=of bern] (one) {1};
    \node[obs, above=of bern, xshift=0cm] (ok_obs) {$y^{(i)}$};
    
    \factoredge {x,nn_params} {h_nn} {y};
    \factoredge {wtheta} {sigma} {y};
    \factoredge {var_w} {nn_params} {};
    \factoredge {} {y} {ok_obs};
    \factoredge {} {y} {trun_obs};
    \factoredge {trun_obs} {bern} {one};

    \hgatep {X-gate} %
  	{(ok_obs)} {$c=0$} %
  	{(trun_obs)(trun_obs-caption)(bern)(bern-caption)(one)} {$c=1$} %
  	{trun} ;
    
    \plate{platew}{
    (nn_params)(var_w)(nn_params-caption)
    }{$K$};
    
    \plate {plate1} { %
    (y)(h_nn)(x)(trun)(ok_obs)(trun_obs)(bern-caption)(X-gate)
  } {$N$}; %
\end{tikzpicture}
\caption{Factor graph describing the devised survival model.
The pseudocode of parameter estimation is shown in Algorithm~\ref{alg:vb_surv}.
The expression 1-CDF$_W$ is the predicted survival function and Bernoulli denotes the Bernoulli distribution. The latent variable $\omega_k$ stands for $\omega_k = [\mu_k, \sigma_k]$. The model predicts the time-to-event from the covariates and the probability of censoring for a given maximum time-to-event.
}
\label{fig:factor}
\end{figure}
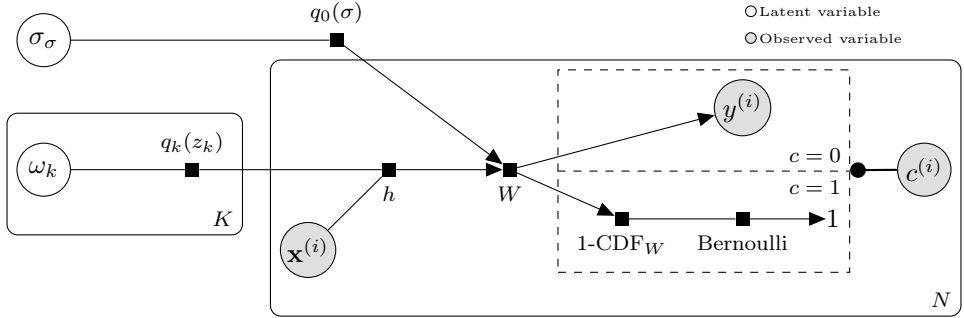

Models relying on the \gls{vb} method are described using the concepts of probabilistic graphical models~\cite{Bishop2006, Koller2009} and directed factor graphs~\cite{minka2008gates}.
The model \eqref{eq:non_lin} is transformed into a \gls{vb} form that takes censoring of the observations into account.
The directed factor graph is shown in \figurename~\ref{fig:factor}. 

The used \gls{vb} approach relies purely on numerical evaluation and requires only proper specification of the model.
That is, only the model likelihood $p(x|\theta)$ and the observations for parameter estimation $x$ have to be specified.
The prior probability distribution $p(\theta)$ is assumed to be constant and the variational family $\mathcal{Q}$ is chosen.
In our case, the likelihood is given by equation \eqref{eq:non_lin} and the variational family  $\mathcal{Q}$ is as described in equation \eqref{eq:meanfield}.

\subsection{Parameter estimation}

As derived by \citet{wingate2013automated}, the gradient of the criterion function is expressed as 
\begin{equation}
\nabla_{\omega} \mathcal{L}(\omega) = \mathbb{E}_{q_{\omega}(\theta)}\left[
\nabla_{\omega} \log q_{\omega}(\theta) \left( \log p(x,\theta) - \log q_{\omega}(\theta)
\right)
\right], 
\label{eq:grad}
\end{equation}
where $\mathcal{L}(\omega)$ is \gls{elbo}. This expression is used in stochastic gradient optimisation to find an estimate of $\omega^*$.


The likelihood of the $i^{\text{th}}$ training data point for a sampled value of $\theta$ is calculated based on equation \eqref{eq:non_lin}.
As illustrated in \figurename~\ref{fig:factor}, the evaluation process depends on whether the observation is censored or not.
The vector of covariates is labelled with $\mathbf{x}^{(i)}$, $y^{(i)}$ is the logarithm of the time-to-event or time-to-censoring, and $c^{(i)}$ is the censoring label.
For a complete observation, i.e. $c^{(i)}=0$, the gradient \eqref{eq:grad} is calculated at the observed $y^{(i)}$.
For censored observations, i.e. $c^{(i)}=1$, the loss is calculated from the predicted probability of censoring at $y^{(i)}$, which equals the survival function 1-CDF$_W$ at $y^{(i)}$.
That is, the predicted probability distribution of $c^{(i)}$ is Bernoulli with the expected value of 1-CDF$_W$.
The complete procedure is also shown as pseudo code in Algorithm~\ref{alg:vb_surv}.
For the initial guess $\omega^{(0)}$, we use $\sigma_{\sigma} = 5$ and $\mu_k = 0$, $\sigma_k = 1\; \forall \; k \in \{1,\ldots,K\}$.

\begin{algorithm}
\begin{algorithmic}[1]

\algnewcommand\algorithmicdraw{\textbf{draw}}
\algnewcommand\Draw[1]{\State\algorithmicdraw\ #1}

\algnewcommand\algorithmicinput{\textbf{Input}}
\algnewcommand\Input[1]{\State\algorithmicinput\ #1}

\algnewcommand\algorithmicoutput{\textbf{Output}}
\algnewcommand\Output[1]{\State\algorithmicoutput\ #1}

\algnewcommand\algorithmicinit{\textbf{Initialize}}
\algnewcommand\Init[1]{\State\algorithmicinit\ #1}


\Input $N$ observations, $\mathbf{x}^{(i)}$ are covariates, $y^{(i)}$ is logarithm of time-to-event or time-to-censoring, and $c^{(i)}$ is censoring label, $i\in\{1,\ldots,N\}$. Variational family $\mathcal{Q}$, prior $p(\theta)$, model structure $p(y|\mathbf{x},\theta)$.
\Output Vector of latent parameters $\omega$, approximating $\omega^*$, specifying the variational density $q_{\omega}(\theta)$
\Init latent parameters $\omega^{(0)}\in\Omega$
\Init rate parameter $\alpha$
\Comment{Using ADAM optimiser, this is adaptable}
\While{ not ELBO convergence }
	\Draw $\sigma \sim q_{0}(\sigma) = \frac{2}{\sigma_{\sigma}}\phi\left(\frac{\sigma}{\sigma_{\sigma}}\right),\,\sigma \geq 0$ \Comment{half-normal distribution}
	\ForAll{$k\in\{1,\ldots,K\}$}\Comment{loop through parameters of the \gls{ann}}
		\Draw $z_k \sim  q_k(z_k) = \mathcal{N}(\mu_k,\sigma_k^2)$
		\Comment{$\theta = [\sigma, \mathbf{z}]$}
	\EndFor 
 	\ForAll{$i\in\{1,\ldots,N\}$} \Comment{loop through observations}
			
		\If {$c^{(i)} = 0$ } \Comment{non-censored observations}
		\State $l^{(i)} = \log p(y^{(i)}| \mathbf{x}^{(i)},\theta) $
\Comment term in stochastic gradient
		\Else \Comment{censored observations}
			\State $\Pr[c^{(i)}|y^{(i)},\mathbf{x}^{(i)},\theta] = 1 - \int_{-\infty}^{y^{(i)}} p(y|\mathbf{x}^{(i)}, \theta)\,dy$
			\Comment probability of surviving beyond $y^{(i)}$ 
			\State $l^{(i)} = \log \Pr[c^{(i)}|y^{(i)},\mathbf{x}^{(i)},\theta]$ 
			\Comment term in stochastic gradient
			
		\EndIf
		
	\EndFor
	\State Compute the gradient $\nabla_{\omega} \mathcal{L}(\omega)$ of equation~\eqref{eq:grad}
	\Comment $\log p(x,\theta) = \sum_{i=1}^N l^{(i)}$, and $q_{\omega}(\theta)$ is known
	\State Update $\omega^{(j+1)} = \omega^{(j)} + \alpha \nabla_{\omega} \mathcal{L}(\omega)$
	
\EndWhile
\end{algorithmic}
\caption{\gls{vb}-based model parameter estimation assuming normal distribution as the variational family.}
\label{alg:vb_surv}
\end{algorithm}

%


\subsection{Estimating the survival function $S(t)$}
Solving the survival model in the Bayesian framework, the survival function $S(t|\mathbf{x})$ for a given value of the covariates $\mathbf{x}$ and of the model latent parameters $\omega$ can be obtained as the posterior prediction distribution using the identified variational distribution $q_{\omega}(\theta)$ instead of the true unknown posterior.
The probability density function is
\begin{equation}
p_{\omega}(t|\mathbf{x}) = \int p(t|\mathbf{x},\theta) q_{\omega}(\theta)\, d\theta.
\label{eq:pred}
\end{equation}
The survival function $S(t|\mathbf{x}) = \Pr[T \ge t]$ is by definition expressed as
\begin{equation}
S(t|\mathbf{x}) = 1 - \int_0^t p_{\omega}(s|\mathbf{x})\,ds = 1 - \int_0^t \!\! \int  p(s|\mathbf{x},\theta) q_{\omega}(\theta)\, d\theta  \,ds
\label{eq:prediint}
\end{equation}
and can be evaluated from~\eqref{eq:prediint} using Monte Carlo integration.
It is a fairly simple and computationally efficient process. The pseudo code is presented in Algorithm~\ref{alg:sx}.

\begin{algorithm}
\begin{algorithmic}[1]
\algnewcommand\algorithmicdraw{\textbf{draw}}
\algnewcommand\Draw[1]{\State\algorithmicdraw\ #1}

\algnewcommand\algorithmicinput{\textbf{Input}}
\algnewcommand\Input[1]{\State\algorithmicinput\ #1}

\algnewcommand\algorithmicoutput{\textbf{Output}}
\algnewcommand\Output[1]{\State\algorithmicoutput\ #1}

\algnewcommand\algorithmicinit{\textbf{Initialize}}
\algnewcommand\Init[1]{\State\algorithmicinit\ #1}

\Input covariates $\mathbf{x}$, 
model structure $p(y|\mathbf{x},\theta)$, variational distribution $\mathcal{Q}$, latent parameters $\omega$, variational model $p_{\omega}(y|\mathbf{x}) = \int  p(y|\mathbf{x},\theta) q_{\omega}(\theta) \, d\theta$, number of samples $N_{\text{MCMC}}$
\Output $S(t|\mathbf{x}) = \Pr(y > \log t|\mathbf{x}) = n(t) / N_{\text{MCMC}}$
\ForAll{$k\in\{1,\ldots,N_{\text{MCMC}}\}$}
\Comment Monte Carlo integration
\Draw $\theta_k \sim q_{\omega}(\theta)$
\Draw $y_k \sim p(y|\mathbf{x},\theta_k)$
\EndFor
\State $b_k(t)=\begin{cases}
    1,& \text{if } y_k > \log t\\
    0,              & \text{otherwise}
\end{cases}$
\State $n(t) = \sum_{k=1}^{N_{\text{MCMC}}} b_k(t)$
\end{algorithmic}
\caption{Prediction of survival function $S(t|\mathbf{x})$.}
\label{alg:sx}
\end{algorithm}

We have mentioned that Monte Carlo integration of the integral~\eqref{eq:evidence} would be too computationally intensive, but the integral~\eqref{eq:prediint} is easy to calculate using the Monte Carlo method even though the integration variable $\theta$ has the same dimensionality in both cases. The reason is that the probability density function $q_{\omega}(\theta)$ is quite different from $p(\theta)$. While $p(\theta)$ is very wide -- we take it to be constant over $\Re^+ \times \Re^K$ -- and the integration would require a lot of samples, the function $q_{\omega}(\theta)$ is much more localized, so every sample contributes a lot more to the accuracy of the result and a much lower number is sufficient.

%
%
%
%
%
%

%% file: results.tex
\section{Variational Bayesian survival analysis applied to unemployment modelling}
\label{sec:pes}

\Gls{pes} records provide a rich description of job seekers in the form of their profiles.
One of the most common metrics that \gls{pes} associates with every job seeker is the so-called probability of exit, referring to ``exiting'' from their records.
It should be noted that not every ``exit'' is due to employment but also events such as retirement, \gls{pes} determining that someone is not genuinely seeking a job, and many others.

Some of the job seekers entering the records at any point in time have exited at a later known date while the others are still unemployed.
From the data perspective, this is a clear example of a right censored data. 
Survival analysis is therefore a viable tool for analysing \gls{pes} data
where the probability of exit from \gls{pes} records serves the function of the hazard rate.

\subsection{Data structure}
Every job seeker is described using 19 covariates.
Some describe personal characteristics of each job seeker such as age, gender, education based on the national classification, last work position based on ESCO classification, municipality of the permanent residence, country of origin, duration of work experience, date of entering the unemployment records, date of employment, and limitations such as disability.
The others describe the interventions of the public employment service, for instance courses attended, employment plan, unemployment benefits, social security benefits and active job seeking grade.
The covariates represent a mix of continuous and discrete variables with very different cardinality.
The complete list of the discrete covariates and their cardinalities is shown in \tablename~\ref{tab:covariates}.
It should be noted that the covariates with cardinality 2 are treated as continuous and do not go through the process of embedding.

\begin{table}[h]
\centering
\caption{List of covariates and their cardinality}
\begin{tabular}{lll}
\toprule
Continuous covariates & \multicolumn{2}{c}{Discrete covariates (cardinality)} \\
\midrule
Day of PES entry & Specific profession category (109) \\
Month of PES entry & Profession program (2336) \\
Age & Municipality (215) \\
Months of work experience & Employment plan status (5) \\
eApplication & PES Office (61) \\
Gender & Reason for PES Entry (10) \\
Employment plan ready & Employability assessment (6) \\
Social benefits & Education category (22) \\
Unemployment benefits & Profession (ESCO) (3772) \\
 & Disabilities (17) \\
\bottomrule 
\end{tabular}
\label{tab:covariates}
\end{table}

\subsection{Model error distribution}
When defining the survival model, one of the key decisions is the selection of the probability density function of the error $W$ in~\eqref{eq:non_lin}.
Since the probability of finding a job decreases over time~\cite{Shimer2012},
 the suitable choices of $W$ are the ones that result in decreasing hazard rate $\lambda(t)$.
The typical choices such as Gumbel or other extreme value distributions behave in the opposite way.
For Gumbel probability density function of $W$, the resulting baseline survival time would follow the Weibull distribution and the hazard rate would rise over time.
Such behaviour is reasonable in many uses of survival analysis but not in modelling of unemployment.

\begin{figure}[h]
\centering
\includegraphics{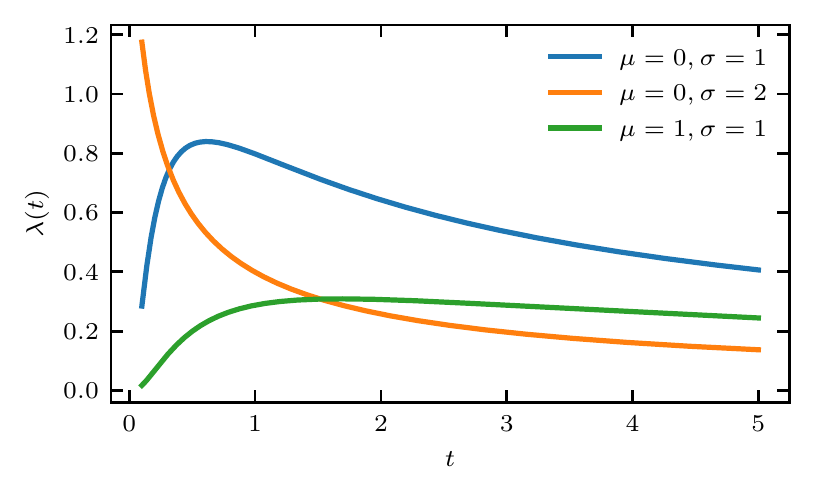}
\caption{Various shapes of the hazard function $\lambda(t)$ for various parameters of the log-normal distribution.}
\label{fig:hazard}
\end{figure}

There are several possible choices of the probability density function of $W$ that result in a time-decreasing hazard rate.
The simplest one is the normal distribution.
For given values of $\mathbf{z}$ and $\mathbf{x}$, the survival time $T$ in~\eqref{eq:non_lin} then follows the log-normal distribution and the hazard rate is monotonically decreasing after its maximum~\cite{Klein2003}.
The probability density function of the event $f(t)$ in~\eqref{eq:prob_t} and the corresponding survival function $S(t)$ become
\begin{equation}
\begin{split}
f(t) &= \frac{\phi\left( \frac{\log t - \mu}{\sigma}  \right)}{t} \quad \textrm{and}\\
S(t) &= 1-\Phi\left( \frac{\log t -\mu}{\sigma} \right),\\
\end{split}
\end{equation}
where $\phi$ and $\Phi$ are the probability density and the cumulative density functions of the standard normal distribution, respectively.
Some typical shapes of the hazard functions for various values of $\mu$ and $\sigma$ are shown in \figurename~\ref{fig:hazard}.

\subsection{Artificial neural network model for the risk function $h_\mathbf{z}(\mathbf{x})$}
\label{sec:nn_structure}
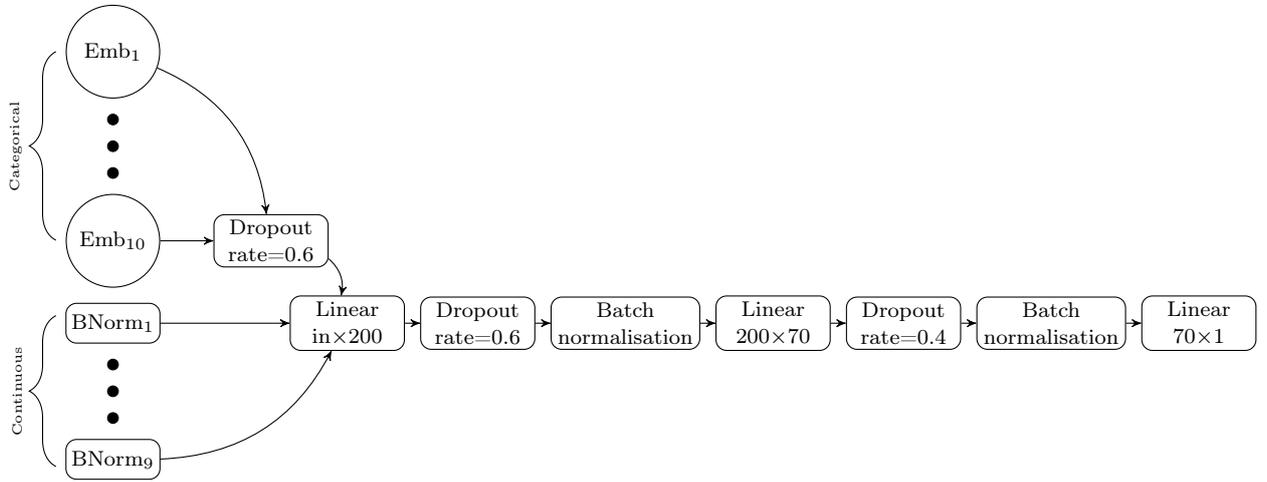
\begin{figure}[h]
\centering
\begin{tikzpicture}
\tikzset{
    >=stealth',
    node distance=.2cm,
    dot/.style={
	circle,draw,fill=black,inner sep=0pt,minimum size=4pt
    },
    punkt/.style={
           circle,
           rounded corners,
           draw=black,
           minimum height=3.5em,
           inner sep=0pt,
           text centered,
           font=\footnotesize,
           execute at begin node=\setlength{\baselineskip}{1.2em}},
    rec/.style={
           rectangle,
           rounded corners,
           draw=black,
           minimum height=1.5em,
           inner sep=2pt,
           text centered,
           font=\footnotesize,
           execute at begin node=\setlength{\baselineskip}{1.2em}},
    lin/.style={
           rectangle,
           rounded corners,
           draw=black,
           text width=9ex,
           minimum height=1.5em,
           inner sep=2pt,
           text centered,
           font=\footnotesize,
           execute at begin node=\setlength{\baselineskip}{1.2em}},        
    bn/.style={
           rectangle,
           rounded corners,
           draw=black,
           text width=12ex,
           minimum height=1.5em,
           inner sep=2pt,
           text centered,
           font=\footnotesize,
           execute at begin node=\setlength{\baselineskip}{1.2em}},
    pil/.style={
           ->,
           thick,
           shorten <=2pt,
           shorten >=2pt,}
}
\node[punkt] (emb0) {Emb$_1$};
\node[dot, below=of emb0] (d1) {};
\node[dot,below=of d1] (d2) {};
\node[dot, below=of d2] (d3) {};
\node[punkt,below=of d3] (emb9) {Emb$_{10}$};
\node[rec,below=of emb9] (c1) {BNorm$_1$};
\node[dot, below=of c1] (cd1) {};
\node[dot,below=of cd1] (cd2) {};
\node[dot, below=of cd2] (cd3) {};
\node[rec,below=of cd3] (c14) {BNorm$_{9}$};

\node[left=of d2,rotate=90,anchor=south,yshift=.8cm] {\tiny Categorical};
\node[left=of cd2,rotate=90,anchor=south,yshift=.8cm] {\tiny Continuous};

\node[lin,right=of emb9,xshift=.5cm] (drop1) {Dropout rate=0.6};
\node[lin,right=of c1,xshift=1.5cm] (lin1) {Linear in$\times$200};
\node[lin,right=of lin1] (d2) {Dropout rate=0.6};
\node[bn,right=of d2] (bn1) {Batch\\normalisation};

\node[lin,right=of bn1] (lin2) {Linear 200$\times$70};
\node[lin,right=of lin2] (d3) {Dropout rate=0.4};
\node[bn,right=of d3] (bn3) {Batch\\normalisation};

\node[lin,right=of bn3] (out) {Linear 70$\times$1};

\path[->] (emb9) edge (drop1);
\path[->] (emb0) edge[bend left] (drop1);
\path[->] (drop1) edge[bend left] (lin1);
\path[->] (c1) edge (lin1);
\path[->] (c14) edge[bend right] (lin1);

\path[->] (lin1) edge (d2);
\path[->] (d2) edge (bn1);
\path[->] (bn1) edge (lin2);

\path[->] (lin2) edge (d3);
\path[->] (d3) edge (bn3);
\path[->] (bn3) edge (out);

\draw [decorate,decoration={brace,amplitude=10pt},xshift=-1.2cm,yshift=-3cm]
(0.45,0.5) -- (0.45,3.0) node [black,midway,xshift=-0.6cm] 
{};

\draw [decorate,decoration={brace,amplitude=10pt},xshift=-1.2cm,yshift=-3cm]
(0.45,-2.5) -- (0.45,-0.5) node [black,midway,xshift=-0.6cm] 
{};

\end{tikzpicture}
\caption{\gls{ann} architecture for describing the risk function $h(\mathbf{x})$.}
\label{fig:nn}
\end{figure}

The underlying risk function $h_\mathbf{z}(\mathbf{x})$ is modelled using a deep \gls{ann} of the architecture shown in \figurename~\ref{fig:nn}.
At the entry level of the network, the dimensions of categorical (discrete) covariates are reduced and the continuous covariates are normalised.
The network then follows three repeating groups of linear layers, normalisation, and dropout~\cite{JMLR:v15:srivastava14a}.
The cardinality of the categorical covariates ranges from simple boolean (yes/no) up to 3772 categories for a covariate describing occupation.
The dimensions are reduced using embedding~\cite{Guo2016} with the reduced number of dimensions equal to
\begin{equation}
n = \min\left(50, \left\lfloor \frac{d}{2} \right\rfloor+1\right),
\label{eq:emb_fun}
\end{equation}
where $d$ is the original number of categories.

\mrev{Comment 1.4}{
As listed in Table~\ref{tab:covariates}, there are 9 continuous covariates and 10 discrete ones.
Using~\eqref{eq:emb_fun}, the categorical values are embedded into 262 dimensions.
Therefore, the first linear node has 271 input and 200 output values.
The structure of the other layers is shown in \figurename~\ref{fig:nn}.
}

\mrev{Comment 1.2.}{
%
The variational distribution of each \gls{ann} parameter $z_i$ is the normal distribution $q_{\omega_i}(z_i)$.
The latent parameters  $\omega_i=\{\mu_i,\sigma_i\}$ represent the mean values and the variances of the variational distribution.
In essence, such a choice means that the resulting variational distribution will not be able to capture any dependencies among the the network parameters.
However, this is not a general limitation of the method.
The mean-field approximation treats the latent parameters of the selected variational distribution as mutually independent~\cite{Blei2017}.
The model parameters, in our case the weights of the neural network, would not have to be treated as mutually independent.
}
The standard normal distribution is used for $W$, and the parameter $\sigma$ is sampled from
a half-normal distribution.

\section{Results}
\label{sec:res}
The data set contains daily updates on every job seeker in Slovenia from 2011 up to 2020.
The network is trained on the data covering 12 months and evaluated on the records from the following 6 months.
The presented results show the evaluation on the first 6 months of 2012 using a network trained on the set spanning the whole year of 2011. 
This is the most dynamic period in the labour market, a consequence of the global financial crisis.
The training set, spanning from January 2011 until December 2011,  contains 99,139 records.
Of those, 55\% have censored events, i.e. a potential exit from \gls{pes} records occurred after the observation window.
The evaluation set, spanning from January 2012 until July 2012, contains 43,641 records.
It should be noted that exit from \gls{pes} records can be due to various reasons, for instance employment, retirement, additional schooling, maternity leave, etc.

\begin{figure}[h]
\centering
\includegraphics{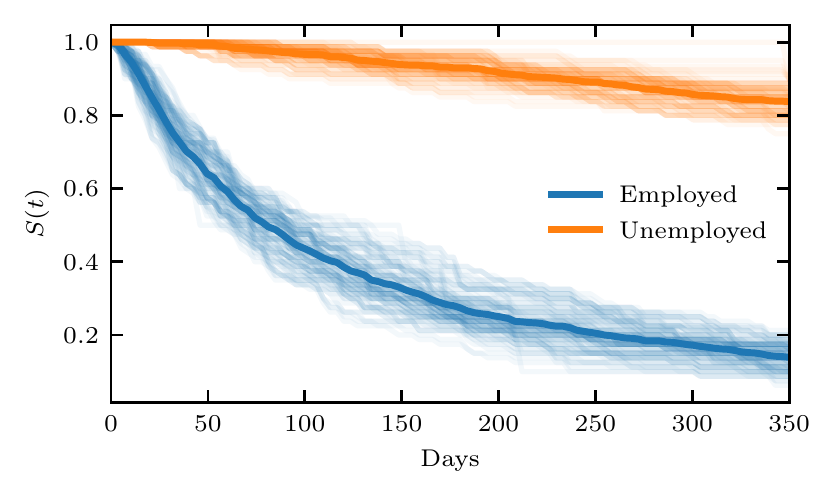}
\caption{Typical shape of the modelled survival functions of the trained model. 
The pale lines are realisations of the $S(t)$ obtained from~\eqref{eq:pred} by sampling the posterior distribution of the model parameters from~\eqref{eq:prediint} using Monte Carlo integration with 200 samples and each bold line is the average of 80 such realisations.}
\label{fig:res_surv}
\end{figure}

\figurename~\ref{fig:res_surv} shows the identified survival curves of two typical job seekers plotted over a period of 1 year starting from the time of entry on the \gls{pes} records.
As expected, the survival probability $S(t)$ for a job seeker unemployed over the next 12 months remains high.
Conversely, for the case of a person that was employed, the survival function drops significantly.
The results on $S(t)$ at a given value of $t$ for the population can be analysed from two viewpoints: pure classification and assessment of the probability of exit.

\subsection{Hyperparameter selection}
\label{sec:hyper}
%

\begin{figure}[h]
\centering
\includegraphics{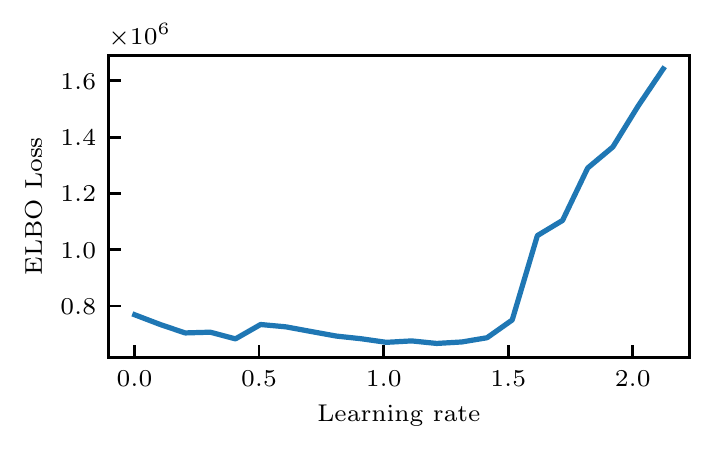}
\caption{Values of \gls{elbo} loss after 4000 iterations for different values of the learning rate parameter.}
\label{fig:lr_test}
\end{figure}

\mrev{Comment 1.5}
{
The optimisation is performed with the ADAM optimiser.
We use two standard approaches for choosing the optimiser's parameters.
Initially, following the guidelines from \citet{smith2018disciplined}, the loss is checked over a range of optimiser's parameters, in particular the learning rate.
Furthermore, the impact of step wise adaptation of the learning rate is also checked as suggested in~\cite{7926641}. 
It should be noted that the whole data set is used in this analysis.

As shown in \figurename~\ref{fig:lr_test}, the minimal value of the loss function over a fixed number of iterations was achieved for learning rate somewhat above 1.
Therefore, 1/10th of this learning rate is used for the training process.
}

\subsection{Stopping criterion}
\label{sec:stopping}
\begin{figure}[h]
\centering
\includegraphics{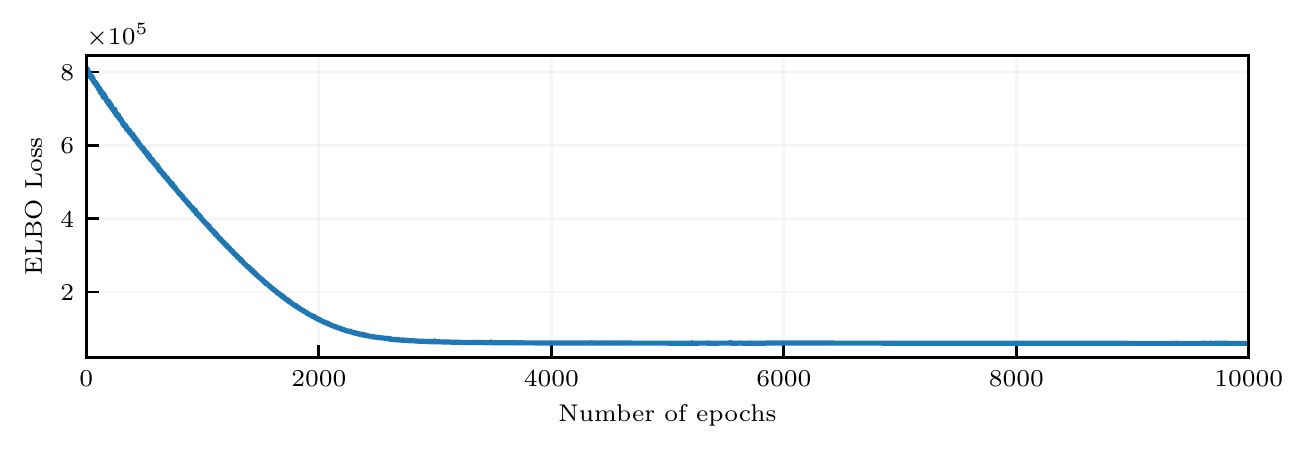}
\caption{Evolution of the \gls{elbo} loss function over number of iterations.}
\label{fig:elbo_converge}
\end{figure}

\mrev{Comment 1.6 - convergence issues}
{
Typically, the stopping criterion is a convergence of the \gls{elbo} loss~\eqref{eq:kl}, i.e., its relative change~\cite{kucukelbir2015automatic}.
There are improved stopping criteria such as~\cite{pmlr-v33-ranganath14,dhaka2020robust}, which use a combination of a smaller step size and Monte Carlo gradient estimates.
The potential of such approaches becomes apparent for high dimensional problems.

In the case analysed here, it turns out that the decrease of the \gls{elbo} value is monotonic and converges to an asymptotic value after a certain number of iterations.
\mrevsec{Comment 2.1}{This is shown in} \figurename~\ref{fig:elbo_converge}.
As a result, in this particular case the stopping criterion can be rather simple, i.e. using the heuristic approach that observes the relative change of the loss function values.
}

\subsection{Classification accuracy}
\begin{figure}[h]
\centering
\includegraphics{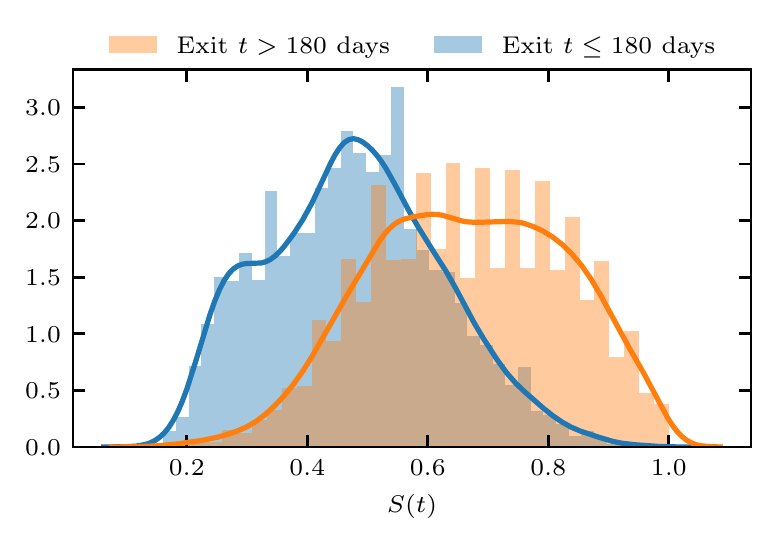}
\caption{Distribution of the survival probability (remaining on the \gls{pes} records) after 180 days for the test data for the period from January to June 2012.
Training was on data from year 2011.}
\label{fig:res_hist}
\end{figure}

Although the overall goal is not classification of the unemployed persons, viewing the results from a classification standpoint offers an easy way for quantifying the performance of the approach.
A simple assessment of the model accuracy can be achieved by analysing the distribution of the survival probabilities of the whole test population at a certain time.
In \figurename~\ref{fig:res_hist}, the values of the survival function $S(t)$ at $t=180$ days from entry are shown separately for the job seekers that exit the records in under 180 days and for the ones that stay unemployed longer.
This value of $t=180$~days is chosen because
more than 180 days of unemployment have a significant negative influence on a job seeker's capability of finding a job~\cite{Shimer2012}.
Therefore, this result can be treated as an indicator of a job seeker's capability of finding a job. 
 
The classification accuracy depends on the selected threshold value. 
For threshold $S(t=180\,\text{days}) =  0.61$, the area under the ROC curve reaches the maximum.
At this value, the classification accuracy is 75.6\%, whereas the trivial model has 50.5\% accuracy.
The trivial model is the case when the whole test set is labelled as either employed or unemployed.
The two groups are shown in~\figurename~\ref{fig:res_hist}.
The blue histogram contains persons that exited \gls{pes} records prior to 180 days, which we label as positive outcome, and the orange histogram are persons that are either still on the records or exited after 180 days.

\subsection{Survival analysis results}
Unlike classification, the estimated survival probability over time $t$ offers a better insight. 
The set of job seekers remaining in the \gls{pes} records for over 180 days is investigated into more detail.
Table~\ref{tab:err} lists the most common outcomes for these job seekers divided into three groups based by the values of $S(t=180\,\text{d})$.
We are particularly interested in the ones for whom the model predicted survival probability  below the threshold $S(t=180\,\text{d})<0.61$.

\begin{table}[h]
\centering
\caption{Detailed overview of job seekers with exit time $T>180$ days based on the value of the survival function $S(t=180\,\text{d})$ and their outcome. The job seekers that exit the records through employment are divided further based on the time of event. The rest tend to end up being not active job seekers, meaning they quit fulfilling their obligations toward \gls{pes}. Most of the remainder are employed by \gls{pes} in public service, and a significant fraction of the rest go on maternity leave.}
\begin{tabular}{lrrrr}
\toprule
& \multicolumn{3}{c}{$S(t=180\,\text{d})$ [\%]}\\
& <0.4 & 0.4--0.61 & >0.61 & Sum [\%]\\
\midrule
Employed, $T\le 240$ days & 1.2 & 6.0 & 2.5 & 9.8\\ 
Employed, $T > 240$ days  & 1.6 & 13.7 & 20.0 & 35.3\\
Not active job seekers & 1.5 & 10.1 & 25.0 & 36.6\\
Public service & 0.1 & 1.6 & 6.2 & 8.0\\
Maternity leave & 0.2 & 1.2 & 1.3 & 2.6\\
Other & 1.5 & 2.3 & 4.0 & 7.7\\
  \midrule
Total & 6.1 & 34.9 & 59.0 & 100.0\\
\bottomrule
\end{tabular}
\label{tab:err}
\end{table}

The false positives -- the job seekers for whom the model  assigns low survival probability $S(t=180\,\text{days})$ but remained for more than 180 days -- deserve further attention as they may not get the necessary \gls{pes} support if the model is relied upon.
In Table~\ref{tab:err}, we split them into two subgroups based on $S(t=180\,\text{d})$ and compare them to the true negatives for whom the model correctly predicts over 180 days to exit.
Let us term employment within a further 60 days, maternity leave, and having a less common outcome lumped as ``other'' a ``good'' outcome.
An underestimate of the date of exit for under 60 days and the failure to take maternity leave into account are inconsequential for \gls{pes} and hopefully for the job seeker.
The ``other'' outcomes are of different desirabilities for the job seeker -- they range from enrolling into further education up to imprisonment -- but they tend to be independent of \gls{pes} actions.
Just like the people employed in 60 extra days and the ones on maternity leave, the ``other'' people are neither failed by \gls{pes} nor a burden of \gls{pes}.
Therefore, assigning low survival probability to such a job seeker will not be likely to cause much harm.

The outcome is ``good'' for 30\% of the false positives, rising to 47\% with the model-predicted $S(t=180\,\text{d})<0.4$.
In contrast, only 13.2\% of the true negatives achieve a ``good'' outcome.
It seems the type II error is more likely for the job seekers for whom it is less damaging, which is fortunate.

The fraction of the long-term job seekers of over 180 days that lose their status and become ``not active'' is the highest in the true negative $S(t=180\,\text{d}) \geq  0.61$ group.
For this group, losing the status is the most likely outcome.
They lose the status by not fulfilling their obligations toward \gls{pes}.
It can be inferred that part of the group are the job seekers that \gls{pes} actions did not help.
Inexplicably from the data, the outcome is very common for job seekers around 60 years of age, indicating that there may be regulatory actions making the status less appealing to them.
Detailed age profile results are shown in \ref{sec:appendix}.

The second most frequent outcome for the true negatives is getting employed after 240 days, thus using \gls{pes} resources and suffering the consequences of unemployment for a longer time.
A lot of the remainder get employed in public service, meaning that they stay in a contractual relationship with \gls{pes}.
The model-predicted $S(t=180\,\text{d})$ is strongly correlated with the probability of the job seeker serving in public service.
This is not surprising as public service is only available to the long-term unemployed and there may be causal connections between the model inputs and the availability of public service to the job seeker.

\section{Discussion on the accuracy}
\label{sec:dis}
Two of the goals of \gls{pes} are to get as many of the job seekers from their records employed as quickly as possible and to lower the number of long-term (over 1 year) unemployed \cite{poslovninacrt}.
The significance of long-term unemployment is that it has various negative consequences, some of which decrease the chance of getting employed \cite{analizaDBO, Shimer2012}.
The presented model could help \gls{pes} better estimate which job seekers need more assistance, potentially improving their service.

However, all of the reported accuracy results on implemented systems addressing unemployment records focus solely on classification.
As a result, not many such systems became operational and others were disbanded due to fears of discrimination of the classification process.
Most recently this happened with the Austrian AMAS system~\cite{Desiere2018}.

Due to the lack of proper profiling model results, the only comparable systems are those that perform classification. 
Even so, comparing the obtained 76\% classification accuracy with the other published results has proven itself challenging.
Due to data privacy regulations, it is not feasible to get access to data sets from various \gls{pes} registries used in the literature.
The algorithms are not published either, so it is not possible to apply them to the data that is available to us.
Model comparison through applying different models to the same data is therefore not possible.
We thus compare the reported results with the results of our model. 
One has to assume that the data sets of various \gls{pes} organisations are of comparable quality and information content, and neglect the differences between labour markets in order to compare various modelling approaches this way.

Results of several systems implemented in \gls{pes} offices are reported by \citet{Scoppetta2018}.
Table~\ref{tab:summary} shows a list of reported models comparable to the presented one, their underlying modelling methods and the resulting accuracies.
In most published cases, the accuracy of the model in predicting the correct probability of exit over time is close to~70\%.

\begin{table*}[h]
\centering
\caption{Summary of reported modelling results addressing the problems of unemployed persons profiling.}
\label{tab:summary}
\begin{tabular}{@{}p{2.7cm}p{7.5cm}r@{}}
\toprule
Reference &  Method & Accuracy \\
\midrule
Australia \cite{Ponomareva2013}  & Logistic regression & Not reported \\
Austria~\cite{Desiere2018} & Logistic regression &80-85\% \\ 
Belgium~\cite{Desiere2018}  & Random forest & 67\% \\
Croatia~\cite{Pojarski2018} & Logistic regression & 69\% \\
Denmark~\cite{Rosholm2004,Madsen2014,Larsen2011} & Logistic regression &66\% \\
Finland \cite{Riipinen2011}  &Statistical model & 89\% \\
France~\cite{10.1111/ijsw.12088}  & Logistic/Random forest/Neural networks &  70\% \\
Ireland \cite{O_CONNELL_2009} &   Probit regression & 69\%\\
Netherlands~\cite{Wijnhoven2014} &  Logistic regression & 70\% \\
New Zealand~\cite{ObbenJames2002Tafp} & Random forest and Gradient boosting & 63-83\% \\ 
\bottomrule\end{tabular}
\end{table*} 

It should be noted that in many cases the reported accuracy listed in Table~\ref{tab:summary} refers to a particular subset of data, limited by gender, location, education, etc.
Furthermore, there is no information on the balance of the test data,
and it is easier to achieve a certain accuracy with a less balanced data set.
The proposed \gls{vb} model is tested on the whole population of job seekers, the data set is balanced (50.5\% remained unemployed for longer than 180 days, and 49.5\% was employed before that threshold), and the achieved accuracy is above the average reported in the literature.
It can therefore be concluded that it works well.
The source code is published and the curves of survival probability for each classification outcome are reported in \figurename~\ref{fig:res_hist}.
It will thus be possible to compare the future models with the presented \gls{vb} algorithm.

%% file: conclusion.tex
\section{Conclusion}
The study successfully introduces advanced survival analysis methods to the question of unemployment.
It demonstrates the use of variational Bayesian methods for performing survival analysis with an arbitrary risk function implemented as an \acrfull{ann}.
The method enables a computationally efficient approximation of the posterior probability distributions. 
It thus becomes possible to exploit the true statistical nature of the survival analysis despite the need of using deep \glspl{ann} with a high number of parameters.

The proposed survival model predicts the time until exit from the job seeker records.
As the most common reason of exiting the records is employment, the model provides time information regarding the probability of employment.
This information is useful both for the job seekers and for the operation of the \acrfull{pes}.
The model can assist the \gls{pes} employment counsellors in recognising the job seekers that do not need \gls{pes} resources as they will get employed soon regardless of the interventions.
It thus enables the counsellors to focus their attention on the job seekers that need help.



Data analysis has been used to make predictions of similar kinds before.
However, the performance of the proposed model cannot be accurately compared with the historical ones because the published results are incomplete.
Based on the available data and the theoretical soundness, we believe that the proposed model performs relatively well.



The resulting model can be used for performing gamification scenarios allowing both the counsellors and the job seekers to explore various strategies for improving the job prospects, i.e. for reducing the survival probability.
A limitation of the study is that it only explores the survival function and not the type of event that results in exiting the unemployment records.
Most exits are of a desirable kind but some are not.
Future research that includes the differences between types of exit and focuses on increasing the likelihood of desirable ones would be highly beneficial.


\section*{Ethics statement}
The data for this analysis was provided by the \glsdesc{pes} in Slovenia in the framework of the EU H2020 HECAT project.
All records were anonymised according to the General Data Protection Regulation of the European Union~\cite{GDPR} and respecting all national privacy laws.

%% file: appendix.tex
\section{Age distribution for inactive persons}
\label{sec:appendix}
\begin{figure}[h]
\centering
\includegraphics{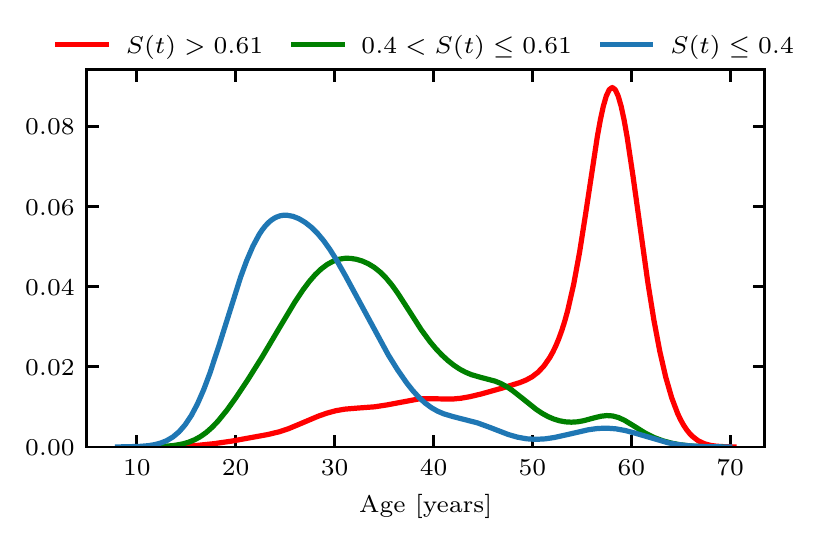}
\caption{Age distribution of persons that were identified by \gls{pes} counsellors as not active job seekers and remained on the \gls{pes} records longer than 180 days.
The age is taken at the moment of entry on the \gls{pes} records.
The plots represent kernel-density estimation of the observed number of records.
}
\label{fig:age}
\end{figure}

The age distribution of unemployed persons that were identified as inactive by the \gls{pes} office is shown in \figurename~\ref{fig:age}.
It is clear that the majority of the cases identified as not active job seekers are persons older than 50 years.
There is no such significant skewness in the age distribution for persons with lower estimated survival probability.
